# Energy scan of correlations in p+p and Be+Be from NA61/SHINE

**A. Seryakov for NA61/SHINE collaboration**
St. Petersburg State University, Russia

E-mail: seryakov@yahoo.com

**Abstract.** The existence of the critical point (CP) of strongly interacting matter is still an open problem. An extensive strong interactions program including a search of the CP and the study of the onset of deconfinement was started by the NA61/SHINE experiment at the CERN SPS. A two dimensional scan of the phase diagram scan is performed to search for the CP and to shed light on the phase transition region. This program includes studies of hadron production in proton-proton, proton-nucleus and nucleus-nucleus interactions measured in a wide range of colliding energy and system size. Correlations between various observables measured at midrapidity as well as in separated rapidity intervals are considered as additional and sensitive tools of this phase diagram scan. We present NA61/SHINE results of studies of energy dependence of two-particle correlations of pseudo-rapidity and azimuthal angle in p+p collisions at the SPS and the first results on correlations between multiplicity and mean transverse momentum in $^7Be + ^9Be$ collisions at 150A GeV/c obtained for separated pseudo-rapidity intervals (so called long-range correlations). Comparison with data calculations using the EPOS 1.99 model are also discussed.

## 1. Introduction

We present preliminary results of two-particles correlation analysis in p+p collisions (previously shown in [1], except figure.3) and pseudo-rapidity correlations in $^7Be + ^9Be$ interactions at 150A GeV/c. Two-particles correlations are an unique tool to disentangle different sources of correlations. Pseudo-rapidity correlations are a powerful method for probing the initial conditions for the formation of the QGP [2][3]. A detailed description of the NA61/SHINE setup can be found in [4].

*1.1. Two particles correlations*
Such correlations were studied as a function of the difference in pseudo-rapidity $\Delta\eta = |\eta1-\eta2|$ and azimuthal angle $\Delta\varphi = |\varphi1- \varphi2|$ between two particles in the same event. The correlation function is calculated as:

$$C(\Delta\eta, \Delta\varphi) = \frac{N_{mixed}^{pairs}}{N_{data}^{pairs}} \frac{S(\Delta\eta, \Delta\varphi)}{M(\Delta\eta, \Delta\varphi)}; \quad S(\Delta\eta, \Delta\varphi) = \frac{d^2N^{signal}}{d\Delta\eta d\Delta\varphi}; \quad M(\Delta\eta, \Delta\varphi) = \frac{d^2N^{mixed}}{d\Delta\eta d\Delta\varphi};$$

where $S(\Delta\eta,\Delta\varphi)$ and $M(\Delta\eta,\Delta\varphi)$ are the distributions for pairs from data and mixed events, respectively. The $\Delta\varphi$ range is folded. Pseudo-rapidity is calculated in the center of mass system with the pion mass assumption for all charged particles, which allows to compare our results with the RHIC and LHC results.

*1.2. Pseudo-rapidity correlations*

Pseudo-rapidity correlations are defined as correlations between observables *B* and *F* in different pseudo-rapidity windows. The correlation parameter is computed as:

$$b_{rel}[B,F] = \frac{\langle F \rangle}{\langle B \rangle} \frac{\langle BF \rangle - \langle B \rangle \langle F \rangle}{\langle F^2 \rangle - \langle F \rangle^2}$$

where <*B*> and <*F*> are the event averages in the «backward» and «forward» pseudo-rapidity windows respectively. In this work we present the following types of pseudo-rapidity correlations:

- Backward multiplicity $N_B$ and forward multiplicity $N_F$: $b_{rel}[N_B, N_F]$
- Backward mean transverse momentum $Pt_B$ and $N_F$: $b_{rel}[Pt_B, N_F]$
- Backward and forward mean transverse momenta $Pt_B$ and $Pt_F$: $b_{rel}[Pt_B, Pt_F]$

## 2. Analysis procedure

All results were corrected for biases due to trigger, off-line selection, track and event selection etc. by using events obtained from EPOS 1.99 model event generator before and after GEANT detector simulation and reconstruction. Electrons and positrons were removed by a dE/dx cut. Possible effects from hard scattering were reduced by a cut on particle transverse momentum $p_T$ < 1,5 GeV/c. All results were calculated inside NA61/SHINE acceptance, the influence of which will be shown in the example of the EPOS 1.99 model in the next chapter. Only statistical errors are shown for pseudo-rapidity correlations.

## 3. Results

*3.1. Two particles correlations*

The correlation function $C(\Delta\eta, \Delta\varphi)$ for p+p collisions of different energies and different charged pair combinations is shown in figure. 1. There are several notable structures: a maximum at $(\Delta\eta,\Delta\varphi)= (0,\pi)$, likely a result of resonance decays and momentum conservation. It is stronger for unlike-sign pairs and significantly weaker for same charge pairs; an enhancement at $(\Delta\eta,\Delta\varphi) = (0,0)$, probably due to Coulomb interactions (unlike-sign pairs) and quantum statistics (same charge pairs). Results of calculations with the EPOS 1.99 model are also shown in figure. 3. The model calculations are consistent with the enhancement seen at $(0,\pi)$ (compare figure. 3 left and right). Moreover, the model the results in the NA61/SHINE acceptance and $4\pi$ acceptance are very similar (compare figure.3 center and right).

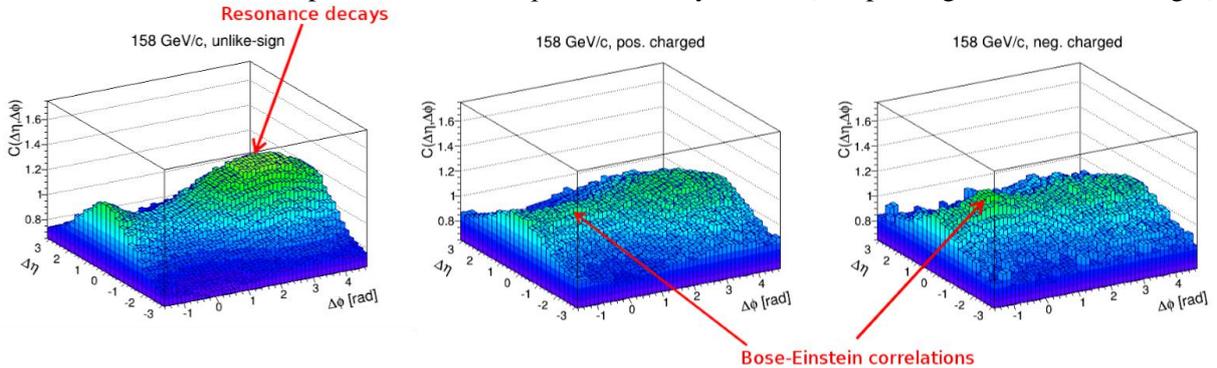

**Figure 1.** Results on $\Delta\eta\Delta\varphi$ correlations for inelastic p+p interactions at 158 GeV/c. Results for unlike-sign (left), pos. charged (middle) and neg. charged (right) pairs are shown. The maximum at $(\Delta\eta,\Delta\varphi) = (0,\pi)$ is probably a result of resonance decays and momentum conservation. The enhancement at $(\Delta\eta,\Delta\varphi) = (0,0)$ is likely due to Coulomb interactions and quantum statistics.

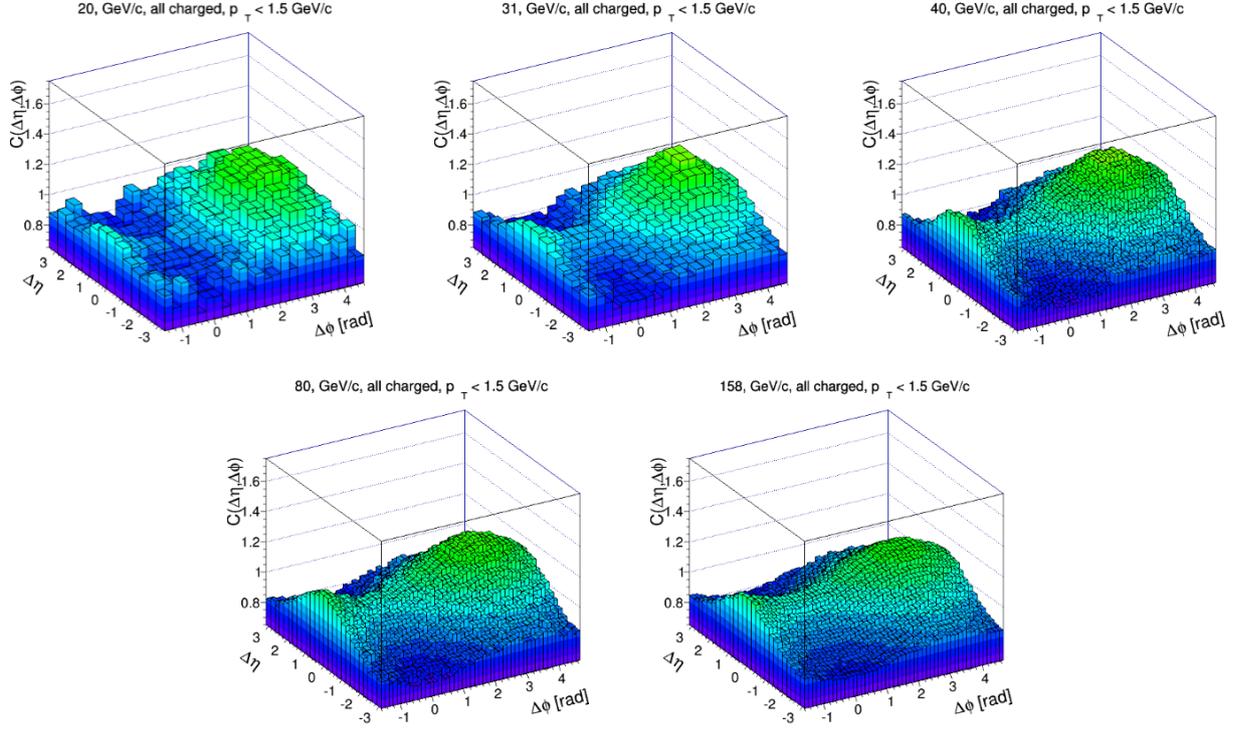

**Figure 2.** Energy dependence of $\Delta\eta\Delta\varphi$ correlations for inelastic p+p interactions.

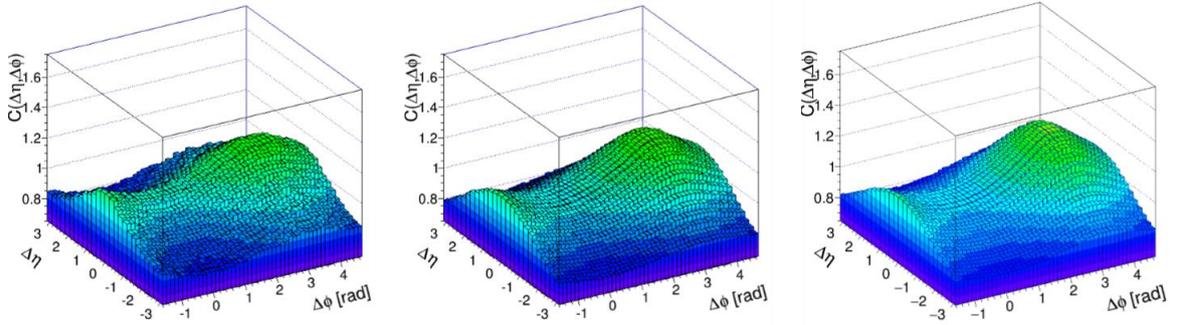

**Figure 3.** $\Delta\eta\Delta\varphi$ correlations for all charged particles in inelastic p+p interactions at 158 GeV/c. NA61/SHINE (left), EPOS 1.99 inside the NA61/SHINE acceptance (middle) and EPOS 1.99 in full $4\pi$ acceptance (right).

*3.2. Pseudo-rapidity correlations*
We used two sets of windows for the study of pseudo-rapidity correlations in $^7$Be+$^9$Be collisions at 150A GeV/c: short-range (figure 4) and with fixed forward window (figure 6). A strong dependence of the correlations on the window position is observed. EPOS 1.99 describes the data only qualitatively.

**4. Conclusions**
Measurements in p+p collisions show structures in two-particle correlations of pseudo-rapidity and azimuthal angle coming mainly from resonance decays, conservation laws, quantum statistics and Coulomb interactions. The enhancement at (0,0) increases, the hill at (0,$\pi$) decreases with energy.

Correlations in pseudo-rapidity of multiplicity and mean transverse momentum in $^7$Be+$^9$Be collisions at 150A GeV/c strongly depend on the position of the chosen pseudo-rapidity windows. Significant

multiplicity–mean transverse momentum long-range correlations were observed (figure 6) that might be a signal of collective effects.

**Acknowledgments**

This work was supported by the Polish National Center for Science (grants 2011/03/N/ST2/03691, 2012/04/M/ST2/00816 and 2013/11/N/ST2/03879), the Federal Agency of Education of the Ministry of Education and Science of the Russian Federation and SPbSU research grant 11.38.193.2014.

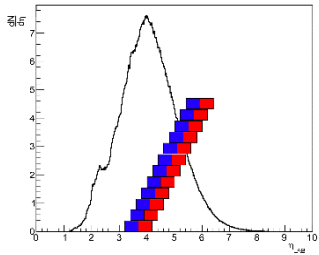

**Figure 4.** Illustration of connected η-windows configurations.

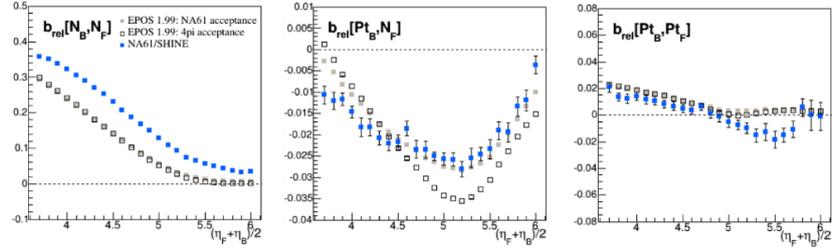

**Figure 5.** Pseudo-rapidity multiplicity and event mean transverse momenta correlation coefficient as a function of windows connection point (see figure 4) for Be7+Be9 150A GeV/c. EPOS 1.99 inside NA61/SHINE acceptance (grey squares) and 4π-acceptance (open squares) are shown for comparison.

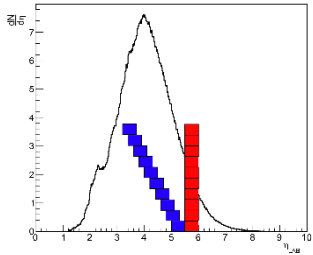

**Figure 6.** Illustration of disconnected η-windows configurations.

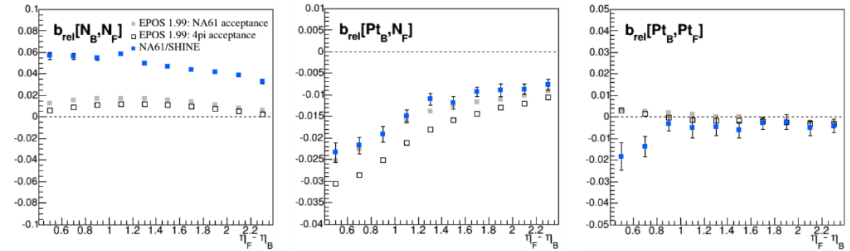

**Figure 7.** Pseudo-rapidity multiplicity and event mean transverse momenta correlation coefficient as a function of distance between windows (see figure 6) for Be7+Be9 150A GeV/c. EPOS 1.99 inside NA61/SHINE acceptance (grey squares) and 4π-acceptance (open squares) are shown for comparison.